\documentclass[10pt]{article}
 
\usepackage{amsfonts, epsfig, amsmath, amssymb, color,mathabx,amsthm}

\newcommand{\rmd}{\mathrm{d}}
\newcommand{\rme}{\mathrm{e}}
\newcommand{\rmi}{\mathrm{i}}

\theoremstyle{definition}

\newtheorem{exa}{Example}[section]

\begin{document}

\title{Marcus versus Stratonovich for Systems with Jump Noise}

\author{Alexei Chechkin\footnote{Kharkov Institute of Physics and Technology, 
Akademicheskaya St.\ 1, 61108 Kharkov, Ukraine  and Max Planck Institute for the Physics of Complex Systems, 
N\"othnitzer Str.\ 38, 01187 Dresden, Germany (\texttt{achechkin@kipt.kharkov.ua}).} \ 
and Ilya Pavlyukevich\footnote{Friedrich Schiller University Jena, Faculty of Mathematics and Computer Science, Institute for Mathematics,
07737 Jena, Germany (\texttt{ilya.pavlyukevich@uni-jena.de}).}}

\date{\null}

\maketitle
\begin{abstract}
The famous It\^o--Stratonovich dilemma arises when one examines a 
dynamical system with a multiplicative white noise. In physics literature, this dilemma is often 
resolved in favour of the Stratonovich prescription because of its two characteristic properties valid for 
systems driven by Brownian motion: 
(i) it allows physicists to treat stochastic integrals 
in the same way as conventional integrals, and 
(ii) it appears naturally as a result of a small correlation time 
limit procedure.
On the other hand, the Marcus prescription [\textit{IEEE Trans.\ Inform.\ Theory} {\bf 24}, 164 (1978);
\textit{Stochastics} {\bf 4}, 223 (1981)] should be used to retain (i) and (ii) 
for systems driven by a Poisson process, L\'evy flights or more general jump processes.
In present communication we present an in-depth comparison of the It\^o, Stratonovich, and Marcus equations 
for systems with multiplicative jump noise.
By the examples of a
real-valued linear system and a complex oscillator 
with noisy frequency (the Kubo--Anderson oscillator)
we compare solutions obtained with the three prescriptions.
\end{abstract}



\section{Introduction}

The It\^o--Stratonovich dilemma is a remarkable issue in the theory of stochastic 
integrals and stochastic differential equations (SDE) with a white Gaussian noise. 
It has been extensively discussed in physics literature; the references 
include basic monographs on statistical physics \cite{Gardiner-04,vanKampen07,Risken89,HorLef84}. 

The famous It\^o formula gives 
the rule how to change variables in the stochastic It\^o integral \cite{Ito-44}. In particular the usual 
integration by parts is not applicable, and the chain rule (also called Newton--Leibniz rule) does not hold in the
It\^o calculus. The It\^o interpretation is preferable, e.g.\ if the SDE is obtained as a continuous time 
limit of a discrete time problem, as it takes place in mathematical finance 
\cite{ContT-04,SetLeh81} or population biology \cite{Turelli77}.
 
Stratonovich \cite{Strato-66} introduced another form of 
stochastic integral which can be treated according to the conventional rules of integration.
Another important property of a Stratonovich equation concerns its 
interpretation as a Wong--Zakai small correlation time limit of solutions of 
differential equations with Gaussian coloured noise \cite{WongZakai-65}. 
Stratonovich prescription is preferable, e.g.\ in physical kinetics \cite{Risken89,WestBLSS-79,BulsaraLSSW-79,vanKampen81}.

In general, since the white noise is a mathematical idealisation of a real dynamics, the choice of prescription is 
not predetermined and may depend on the dynamical properties of the particular system. 
Thus, Kupferman at el.\ \cite{KupfermanPS-04} showed that an adiabatic elimination procedure in 
a system with inertia and coloured multiplicative noise 
leads to either an It\^o or a Stratonovich equation, depending on whether the noise correlation time 
tends to zero faster or slower than the particle relaxation time. 
We refer the reader to a recent review \cite{ManMcC12} for historical 
background and discussions of some contemporary contributions, and mention the work \cite{KurMiy14} as the newest 
evidence
of the continuing interest to this classical problem.

Until recently, the It\^o--Stratonovich dilemma was discussed in the context of Brownian motion. 
Meanwhile, stochastic systems with multiplicative jump noises also attract increasing attention.
They include systems driven by a Poisson process,  
L\'evy flights, or general L\'evy processes 
\cite{Haenggi-80,Grigoriu-98,BlaBue-00,Schoutens-03,ContT-04,GriSam-04,Pawley-06,ErZhuIuKou-09,GPSS-10,SRZDMP-10,SPRM-11,Srokowski10}. 
However, it is not well-known among physicists that both remarkable properties of
the Stratonovich integral are violated if the driving process has jumps. 
In \cite{Marcus-78,Marcus-81} S.\ Marcus fixed this problem by introducing an SDE 
of a new type, whose solution pertains the 
features incident to the Stratonovich calculus in the continuous case. 
Although a Marcus equation 
(also called canonical equation) has been well treated in mathematical literature \cite{KurtzPP-95,Kunita-04,Applebaum-09},
there is only a very few papers in physics literature, which discuss this issue. 
Motivated by the investigation of stochastic energetics for jump processes, Kanazawa et al.~\cite{KanazawaSH-12} essentially followed the 
Wong--Zakai smoothing approach to define an SDE driven by
a multiplicative white jump noise. They eventually re-derived a Marcus canonical equation and then applied 
it to study heat conduction by non-Gaussian noises from two athermal 
environments \cite{KanazawaSH-13}.  
Li et al.\ in \cite{LiMinWang-13,LiMinWang-14err} gave an introduction to Marcus calculus via two equivalent constructions 
used in mathematical and engineering
literature \cite{Marcus-78, Marcus-81, DiPaolaF-93,DiPaolaF-93JAM,SunDuanLi-13} 
and developed a path-wise simulation algorithm allowing to compute thermodynamic quantities.
Further, in \cite{LiMinWang-14} Li et al.\  extended the approach by Kupferman et al.\ \cite{KupfermanPS-04} to the case of a Poisson 
coloured noise. Similarly to \cite{KupfermanPS-04}, in certain parameter regimes they obtained either It\^o or Marcus canonical equations.  

In this paper we present an in-depth comparison of the It\^o, Stratonovich, and Marcus equations 
for systems driven by jump noise. 
In order to preserve the Markovian nature of solutions, we consider coloured noise being an Ornstein--Uhlenbeck
process driven by a Brownian motion or a general L\'evy process.

In the pure Brownian case, we recover the Stratonovich equation 
as a small relaxation time limit of differential equations driven by the Ornstein--Uhlenbeck process. Although 
the passage to the white noise limit should not depend on the smoothing procedure, in the jump case 
we give an instructive derivation of the Marcus equation as a limit of the Ornstein--Uhlenbeck 
coloured jump noise approximations.

We analyse the SDEs in the It\^o, Stratonovich and Marcus form for
two generic examples, namely for a real-valued linear system with multiplicative white noise and a complex oscillator 
with noisy frequency (the Kubo--Anderson oscillator). 
In case of the Kubo--Anderson oscillator we discover a remarkable similarity of
solutions to the Stratonovich and Marcus equations. Nonetheless, from the physical point of view, the Marcus equation 
seems to be a more consistent and natural tool for description of a physical system with bursty dynamics 
or subject to jump noise.

\section{It\^o and Stratonovich Calculus for Brownian Motion}

The definitions of the It\^o and Stratonovich integrals w.r.t.\ the Brownian motion $W$ are well known. 
For a non-anticipating stochastic process $Y$ we define the It\^o integral as a limit
\begin{equation}
\label{eq:iw}
\int_0^t Y_s\,\rmd  W_s:=\lim_{n\to\infty} \sum_{k=1}^n Y_{t_{k-1}} (W_{t_k}-W_{t_{k-1}})
\end{equation}
and the Stratonovich integral as  
\begin{equation}
\label{eq:sw}
\int_0^t Y_s\circ\rmd  W_s:=\lim_{n\to\infty}\sum_{k=1}^n \frac{Y_{t_k}+Y_{t_{k-1}}}{2} (W_{t_k}-W_{t_{k-1}}),
\end{equation}
where $0=t_0<\cdots<t_n=t$ is a partition with the vanishing mesh $\max_{0\leq k\leq n}|t_k-t_{k-1}|\to 0$ as $n\to\infty$.
We refer the reader to  \cite[Chapter 4.2]{Gardiner-04} for a discussion about the mathematical properties and 
physical interpretations of these objects. We recall here two simple examples of stochastic integrals.
 
\begin{exa}
A straightforward calculation based on the definitions \eqref{eq:iw} and \eqref{eq:sw} yields:
\begin{align}
&\int_0^t W_s\,\rmd  W_s =\frac{W_t^2}{2}-\frac{t}{2},  &&\int_0^t W^2_s\, \rmd  W_s=\frac{W^3_t}{3}-\int_0^t W_s\,\rmd s,\\
&\int_0^t W_s\circ \rmd  W_s=\frac{W_t^2}{2},           &&\int_0^t W^2_s\circ \rmd  W_s=\frac{W^3_t}{3}.
\end{align}
As we see, the Stratonovich calculus pertains the Newton--Leibniz integration rule. 
\end{exa}

Consider now the It\^o and Stratonovich SDEs with multiplicative noise, see \cite[Chapter 4.3]{Gardiner-04}:
\begin{equation}
\label{eq:sdeito}
X_t=x+\int_0^t a(X_s)\,\rmd  s+ \int_0^t b(X_s)\, \rmd  W_s
\end{equation}
and 
\begin{equation}
\label{eq:sdestrat} 
X_t^\circ =x+\int_0^t a(X_s^\circ)\,\rmd  s+ \int_0^t b(X_s^\circ)\circ \rmd  W_s.
\end{equation}
It is well known that the Stratonovich equation can be rewritten in the It\^o form as
\begin{eqnarray}
X_t^\circ =x+\int_0^t \Big( a(X_s^\circ)+\frac{1}{2}b'(X^\circ_s)b(X^\circ_s)\Big)\,\rmd  s+ \int_0^t b(X_s^\circ)\, \rmd  W_s.
\end{eqnarray}
For a twice differentiable function $F$, the chain rules for the solutions of these equations read
\begin{align}
\label{eq:ito}
F(X_t)&=F(x)+\int_0^t F'(X_s)\,\rmd  X_s+ \frac{1}{2}\int_0^t F''(X_s) b^2(X_s)\, \rmd  s,\\ 
\label{eq:strat}
F(X_t^\circ)&=F(x)+\int_0^t F'(X_s^\circ)\circ\rmd  X_s.
\end{align}
With the help of Eqs.\ \eqref{eq:ito} and \eqref{eq:strat} we solve two simple linear stochastic differential equations. 
\begin{exa}
The equations for the real-valued linear system with multiplicative noise in the It\^o and Stratonovich form, see \cite[\S 4.4.2]{Gardiner-04},
read 
\begin{equation}
X_t=1+\int_0^t X_{s}\, \rmd  W_s\quad \mbox{and}\quad X_t^\circ=1+\int_0^t X_{s}^\circ\, \circ \rmd  W_s
\end{equation}
and have unique solutions
\begin{equation}
X_t=\rme^{W_t-\frac{t}{2}} \quad \mbox{and}\quad X_t^\circ=\rme^{W_t} ,
\end{equation}
respectively.
\end{exa}

\begin{exa}
\label{ex:strat1}
The Kubo--Anderson oscillator with noisy frequency, see \cite[\S 4.4.3]{Gardiner-04}, is described by a 
complex-valued SDE driven by a Brownian motion with linear drift. Let $w_t=\omega_0 t +\sigma W_t$, $\sigma^2>0$ being a noise variance, and 
$\omega_0\in\mathbb R$ a constant frequency.
Consider an SDE in the sense of It\^o and Stratonovich:
\begin{equation}
Z_t=Z_0+\rmi \int_0^t Z_{s}\, \rmd  w_s\quad \mbox{and}\quad Z_t^\circ=Z_0+\rmi \int_0^t Z_{s}^\circ\circ\rmd  w_s
\end{equation}
It is easy to check with the help of Eqs.\ \eqref{eq:ito} and \eqref{eq:strat}, that the solutions to these equations are
\begin{equation}
\label{eq:expito}
Z_t=Z_0\rme^{\frac{\sigma^2}{2}t}\rme^{\rmi (\omega_0 t +\sigma W_t)}\quad \mbox{and}\quad 
Z_t^\circ=Z_0 \rme^{\rmi(\omega_0 t +\sigma W_t)}.
\end{equation}
It is seen from Eq.\ \eqref{eq:expito} that the It\^o solution has an exponentially increasing amplitude and is not physically relevant.
\end{exa}

\bigskip

\begin{figure}
\begin{center}
\includegraphics[width=.6\textwidth]{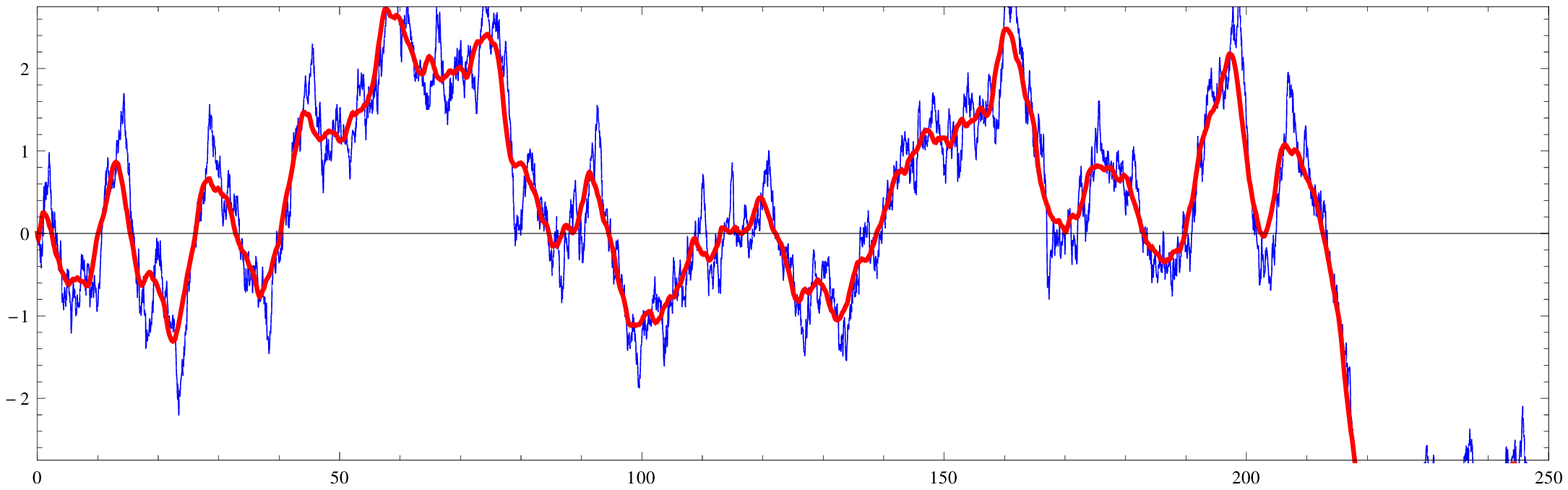}

\includegraphics[width=.6\textwidth]{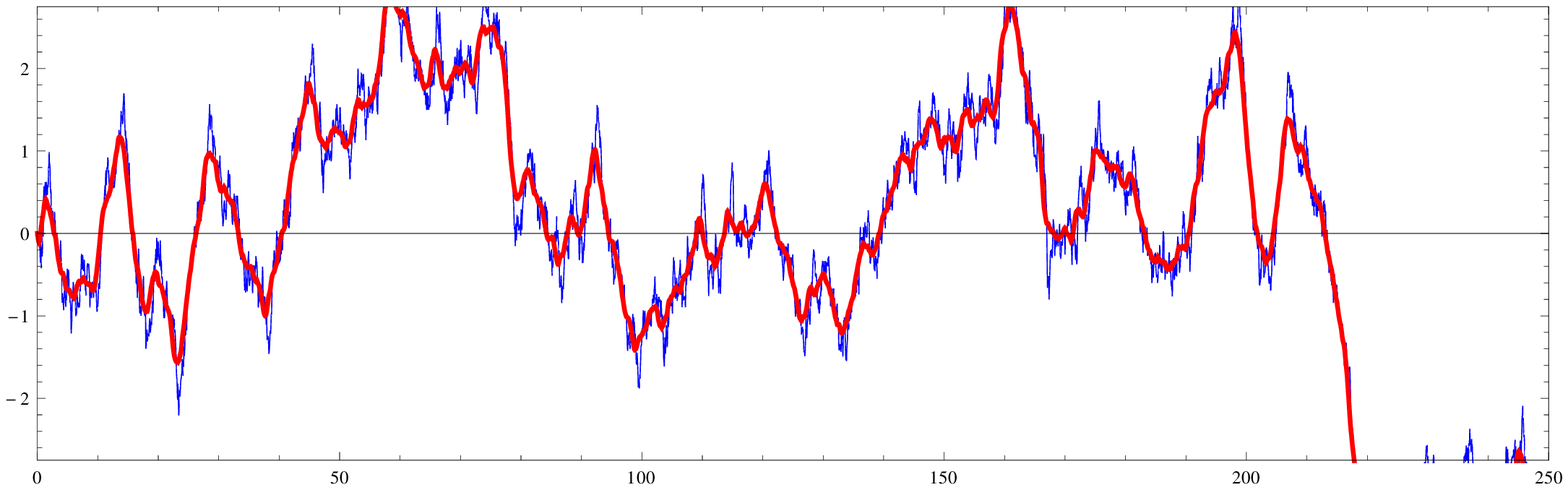}

\includegraphics[width=.6\textwidth]{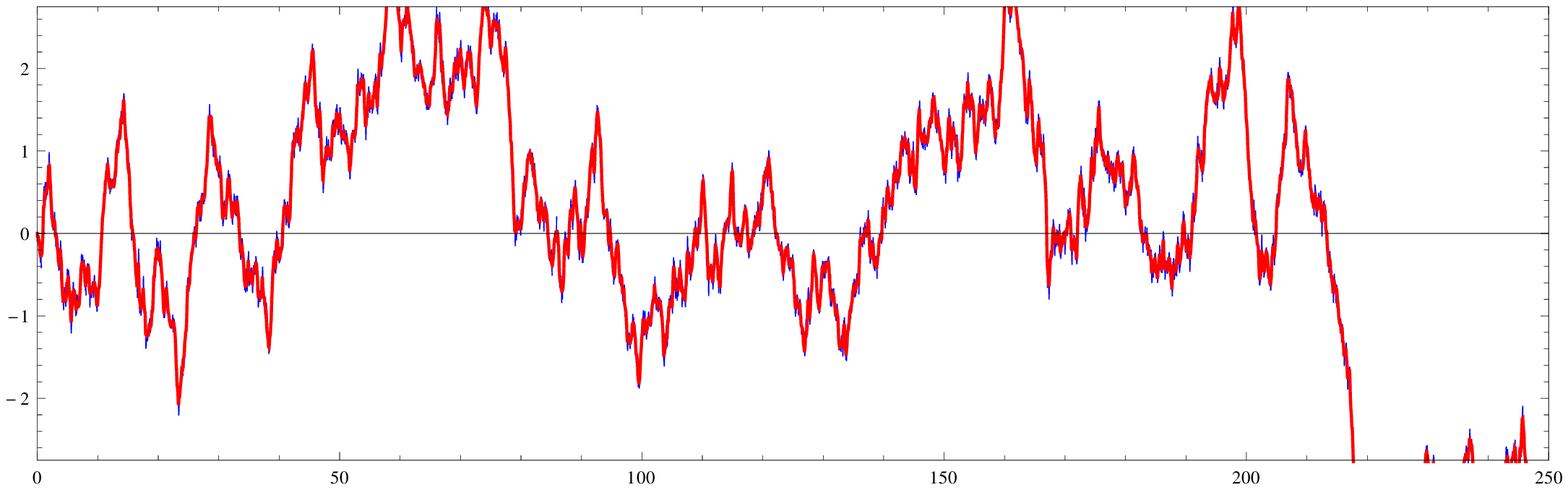}
\end{center}
\caption{(Colour online). Coloured approximations $W^\tau$ (red) of the Brownian motion $W$ (blue) with decreasing 
relaxation times $\tau=20, 1$, and $0.1$ (from top to bottom).
\label{f:gauss}}
\end{figure}

Along with the Newton--Leibniz rule \eqref{eq:strat},
another important feature of a Stra\-to\-no\-vich SDE is that it can be considered as a limit 
of differential equations driven by smooth approximations of the Brownian motion. This interpretation goes back to Wong and Zakai 
\cite{WongZakai-65}. Instead of polygonal smoothing 
usually used in the literature \cite{Arnold-74}, we employ coloured noise approximations $\dot W^\tau$ in the form of 
the Ornstein--Uhlenbeck process with small correlation time $\tau$. They are obtained from the Langevin equation
\begin{equation}
\ddot W_t^\tau=-\frac{1}{\tau} \dot W^\tau+\frac{1}{\tau}\dot W,
\end{equation}
which is solved explicitly as
\begin{equation}
W^\tau_t=\int_0^t (1-\rme^{-\frac{t-s}{\tau}})\,\rmd  W_s\quad \mbox{and} \quad 
\dot W^\tau_t=\frac{1}{\tau} \int_0^t \rme^{-\frac{t-s}{\tau}}\,\rmd  W_s.
\end{equation}
It can be easily shown that $W^\tau$ tends to $W$ as $\tau\to 0$ (see Fig.~\ref{f:gauss}), so that $\dot W^\tau$ can 
be seen as $\tau$-correlated approximations of the
delta-correlated white noise $\dot W$.
Let us now substitute $W$ in Eq.\ \eqref{eq:sdeito} by its approximation $W^\tau$ and consider 
a $\tau$-dependent differential equation 
\begin{equation}
 X^\tau_t=x+\int_0^t a(X^\tau_s)\, \rmd  s+\int_0^t b(X_s^\tau)\dot W^\tau_s\, \rmd s.
\end{equation}
Assume that $b(x)>0$ and define  a function $F(x)=\int_0^x \frac{\rmd  y}{b(y)}$ which is strictly monotone and smooth. Then the 
Newton--Leibniz rule of the conventional calculus gives 
\begin{equation}
F(X^\tau_t)=F(x)+\int_0^t F'(X^\tau_s)\,\rmd  X^\tau_s
=F(x)+\int_0^t \frac{\displaystyle a(X^\tau_s)}{\displaystyle b(X^\tau_s)}\,\rmd  s+W^\tau_t.
\end{equation}
Since $W^\tau$ converges to $W$ as $\tau \to 0$, $X^\tau$ converges to a limit $X^\circ$ which satisfies the equation
\begin{equation}
F(X^\circ_t)=F(x)+\int_0^t \frac{\displaystyle a(X^\circ_s)}{\displaystyle b(X^\circ_s)}\,\rmd  s+W_t.
\end{equation}
Let $G$ denote the inverse of $F$, that is $G(F(x))=x$. Taking into account that 
$G'(x)=b(x)$ and $G''(x)=b'(x)$, we apply the formula \eqref{eq:ito} with the function $G$ to the solution $\displaystyle F(X^\circ_t)$
to obtain the equality 
\begin{equation}
\begin{aligned}
X_t^\circ= G(F(X_t^\circ))
&=x+\int_0^t a(X_s)\, \rmd  s+\int_0^t b(X_s^\circ)\, \rmd  W_s+\frac{1}{2}\int_0^t b'(X_s^\circ)b(X_s^\circ)\, \rmd  s\\
&=x+\int_0^t a(X_s^\circ)\, \rmd  s+\int_0^t b(X_s^\circ)\circ \rmd  W_s.
\end{aligned}
\end{equation}
Hence, the process $\displaystyle X^\circ$ solves the Stratonovich SDE \eqref{eq:sdestrat}.

\section{It\^o and Stratonovich calculuses for processes with jumps\label{s:is}}

With the help of formulae \eqref{eq:iw} and  \eqref{eq:sw} one can also define It\^o and Stratonovich integrals for a broader class
of processes with jumps, in particular for L\'evy processes 
and for semimartingales, see \cite[Chapter 1]{Kunita-04}.

For simplicity we restrict ourselves to integrals and SDEs driven by a L\'evy process with finite number of jumps, which is a sum
of a Brownian motion with drift and an independent compound Poisson process.
Let $P=(P_t)_{t\geq 0}$ be a Poisson process with intensity $\lambda>0$ and arrival times $T_0=0$, $(T_m)_{m\geq 1}$, 
such that the waiting times $T_m-T_{m-1}$ are i.i.d.\ exponentially distributed with the mean $\lambda^{-1}$.
 Let
\begin{equation}
\label{eq:cpp}
N_t=\sum_{m=1}^{P_t} J_m =\sum_{m=1}^\infty J_m\mathbb{I}_{[T_m,\infty)}(t)
\end{equation}
be a compound Poisson process with the i.i.d.\ jumps $(J_m)_{m\geq 1}$ being independent of $P$. 
Here $\mathbb{I}_{[T,\infty)}(t)$ is the indicator function being $1$ on $[T,\infty)$ and $0$ otherwise.
For $\omega_0\in \mathbb R$, $\sigma\geq 0$ 
and a 
Brownian motion $W$ denote $L_t=\omega_0 t+\sigma W_t+N_t$.

For a trajectory of a random process $Y$ we denote by $\Delta Y_t$ the jump size of $Y$ at the time instant $t$, i.e.\
$\Delta Y_t=Y_t-Y_{t-}$, where $Y_{t-}=\lim_{h\downarrow 0} Y_{t-h}$.

To compare the continuous and jump calculuses, we give a couple of basic integration examples.
\begin{exa}
Let $P$ be a Poisson process. Then the integration in the It\^o sense \eqref{eq:iw} gives
\begin{equation}
\begin{aligned}
\int_0^t P_s\,\rmd  P_s&=\sum_{s\leq t}P_{s-}\Delta P_s=\frac{P_t^2}{2}-\frac{P_t}{2}\\
\int_0^t P_s^2\,\rmd  P_s&=\sum_{s\leq t}P_{s-}^2\Delta P_s=\frac{P_t^3}{3}-\frac{P_t^2}{2}+\frac{P_t}{6}.
\end{aligned}
\end{equation}
whereas the integration in the Stratonovich sense \eqref{eq:sw} yields
\begin{equation}
\int_0^t P_s\circ\rmd  P_s=\frac{P_t^2}{2}\quad \mbox{ and }\quad 
\int_0^t P_s^2\circ\rmd  P_s=\frac{P_t^3}{3}+\frac{P_t}{2}.
\end{equation}
\end{exa}
As we see, even in the simple case of a squared Poisson process as an integrand, 
the Stratonovich calculus does not obey the Newton--Leibniz integration rule\footnote{Note that in \cite{Grigoriu-98}, the
so--called Fisk--Stratonovich definition of the Stratonovich integral for jump processes is used. 
It is different from \eqref{eq:sw} and leads to a trivial equivalence between the It\^o
and Stratonovich calculuses in the pure jump Poissonian case.}.

Similarly to the previous section, we consider It\^o and Stratonovich SDEs, see Eqs.\ \eqref{eq:sdeito} and \eqref{eq:sdestrat} with a jump process
instead of a Brownian motion. 

\begin{exa}
\label{ex:expjump}
We solve the equations for the real-valued linear system with the multiplicative Poisson noise in the It\^o and Stratonovich form
\begin{equation}
X_t=1+\int_0^t X_{s}\, \rmd  ( zP_s)\quad \mbox{and}\quad X_t^\circ=1+\int_0^t X_{s}^\circ\, \circ \rmd  ( zP_s),
\end{equation}
where $z\in\mathbb{R}$ is a jump size. 
The solution of the It\^o equation is the so-called stochastic exponent and the solution exists and is unique for $z>-1$:
\begin{equation}
\label{eq:eito}
 X_t=\prod_{s\leq t}(1+z\Delta P_s)=(1+z)^{P_t}.
\end{equation}
To solve the Stratonovich equation, we note that at the arrival time $T_m$ the solution satisfies the equality
\begin{equation}
X_{T_m}^\circ=X_{T_m-}^\circ+\frac{X_{T_m-}^\circ +X_{T_m}^\circ}{2} z.
\end{equation}
This yields for $z\neq 2$
\begin{equation}
X_{T_m}^\circ=
\frac{2+z}{2-z}X_{T_m-}^\circ
\end{equation}
Consequently, the  solution of the Stratonovich SDE is found in the form
\begin{equation}
\label{eq:estrat}
X_t^\circ= \Big(\frac{2+z}{2-z}\Big)^{P_t}.
\end{equation}
\end{exa}

\begin{exa}
\label{ex:kubostrat}
Consider the Kubo--Anderson oscillator perturbed by a centred L\'evy process
$\sigma W_t+z(P_t-\lambda t)$, $\langle \sigma W_t+z(P_t-\lambda t)\rangle=0$.
Denote $l_t=\omega_0 t+\sigma W_t+z(P_t-\lambda t)$ and solve two complex-valued SDEs
in the It\^o and Stratonovich form:
\begin{equation}
Z_t=Z_0+\rmi \int_0^t Z_{s}\,\rmd  l_s\quad \mbox{and}\quad
Z_t^\circ=Z_0+\rmi \int_0^t Z_{s}^\circ\,\circ \rmd  l_s.
\end{equation}
Let us first solve the It\^o equation.
Between the arrival times of Poisson process $P$, the solution of the It\^o equation coincides with the continuous It\^o solution 
\eqref{eq:expito}.
At the arrival time $T_m$ the position of the solution is found from the relation
\begin{equation}
Z_{T_m}=Z_{T_m-} +\rmi Z_{T_m-} z\quad\mbox{and thus}\quad Z_{T_m}=(1+\rmi z)Z_{T_m-}.
\end{equation}
Combining the continuous and the jump parts of the solution and taking into account that 
\begin{equation}
 1+\rmi z=(1+z^2)^{1/2}\rme^{\rmi \varphi(z)}, \quad\mbox{where}\quad 
 \varphi(z)=\arctan z\in\Big(-\frac{\pi}{2},\frac{\pi}{2}\Big) \mbox{ for }z\in\mathbb R,
\end{equation}
we finally obtain a physically inappropriate It\^o solution with exponentially increasing amplitude
\begin{equation}
Z_t=Z_0(1+z^2)^{\frac{P_t}{2}} \rme^{\frac{\sigma^2}{2}t}\rme^{\rmi((\omega_0-\lambda z) t+\sigma W_t+\varphi(z)P_t)}. 
\end{equation}
In the Stratonovich case, as in the Example \ref{ex:expjump}, at the arrival times of $P$ the jumps of $Z^\circ$ satisfy
\begin{equation}
Z_{T_m}^\circ= \frac{2+\rmi z}{2-\rmi z}Z_{T_m-}^\circ
\end{equation}
whereas between the jumps the solution follows the continuous Stratonovich dynamics considered in Example \ref{ex:strat1}.
Noting that
\begin{equation}
 \frac{2+\rmi z}{2-\rmi z}=\rme^{\rmi \psi (z)}, \quad\mbox{where}
 \quad \psi(z)=\arcsin \frac{z}{1+\frac{z^2}{4}}\in\Big(-\frac{\pi}{2},\frac{\pi}{2} \Big) \mbox{ for }z\in\mathbb R,
\end{equation}
we obtain a physically meaningful solution representing stochastic oscillations
\begin{equation}
Z_t^\circ=Z_0\rme^{\rmi ((\omega_0-\lambda z) t +\sigma W_t + \psi(z)P_t)}.
\end{equation}
with constant amplitude.
\end{exa}
In this case it could be instructive to determine the oscillator's line shape. 
Assume that $|Z_0|=1$ and determine the relaxation function
\begin{equation}
\begin{aligned}
\Phi^\circ(t)&=\langle \overline{Z^\circ_0} Z_t^\circ\rangle 
= \langle\rme^{\rmi(\omega_0-\lambda z) t +\sigma W_t + \psi(z)P_t)}\rangle\\
&=\rme^{\rmi (\omega_0-\lambda z) t}\langle\rme^{\rmi \sigma W_t}\rangle\langle\rme^{\rmi \psi(z)P_t}\rangle
=\rme^{\rmi(\omega_0-\lambda z) t}\rme^{-t\frac{\sigma^2}{2}}  \rme^{\lambda t(\rme^{\rmi \psi(z)}-1)}\\
&=\rme^{-t(\frac{\sigma^2}{2} +\lambda(1-\cos \psi(z)))}\rme^{\rmi t (\omega_0+\lambda\sin \psi(z)-\lambda z)}=
\rme^{-\gamma^\circ t+\rmi (\omega_0+\omega^\circ) t},
\end{aligned}
\end{equation}
where
\begin{equation}
\begin{aligned}
\gamma^\circ&=\frac{\sigma^2}{2} +\lambda(1-\cos \psi(z)) ,\\
\omega^\circ&= \lambda(\sin \psi(z)- z).
\end{aligned}
\end{equation}
Then the line shape (see Kubo \cite{Kubo-63}, Eqs.\ (2.6) and (3.6)) has the Lorenzian form 
\begin{equation}
\label{ex:shapestra}
I^\circ(\omega)=\frac{1}{\pi}\mbox{Re}\int_0^\infty \rme^{-\rmi \omega t}\Phi^\circ(t)\,\rmd  t= 
\frac{1}{\pi}\frac{\gamma^\circ}{(\gamma^\circ)^2 +(\omega-\omega_0-\omega^\circ)^2}.
\end{equation}

\section{Coloured jump noise and Marcus SDEs}

As we demonstrated in the previous section, the Stratonovich calculus in the jump case does not pertain 
the Newton--Leibniz change of variables rule. Now we study if it is consistent with the small correlation limit of the coloured noise 
approximations.

\begin{figure}
\begin{center}
\includegraphics{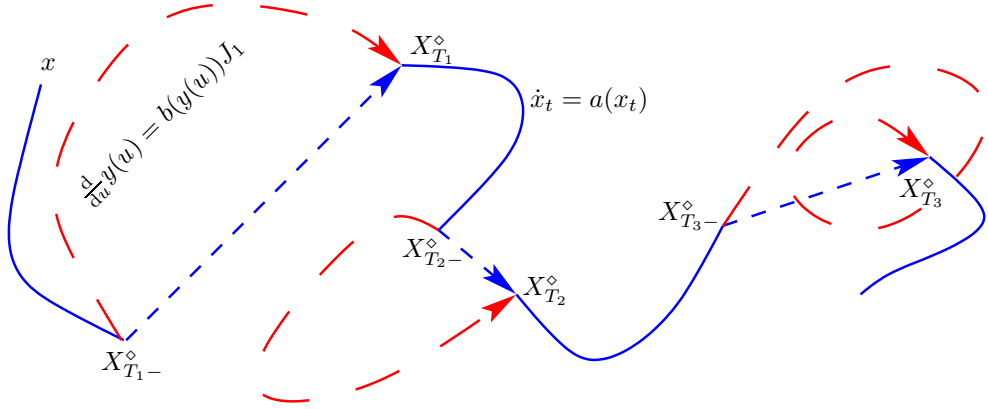} 
\end{center}
\caption{(Colour online) Schematic representation of a solution of Marcus SDE $\rmd  X^\diamond=a(X^\diamond)\,\rmd t+ b(X^\diamond)\diamond \rmd  N$ 
driven by the jump noise
$N_t=\sum_{m=1}^{P_t} J_m$. 
The solution $X^\diamond$ is a jump process with jumps $X_{T_m}^\diamond-X_{T_m-}^\diamond$ (dashed blue lines). 
These jumps are obtained as a 
small relaxation time limit of fast curvilinear motions along the solutions of an auxiliary non-linear ODE
$\frac{\rmd}{\rmd u} y(u)= b(y(u))J_m$ with the initial value $y(0)= X_{T_m-}^\diamond$ (dashed red lines). 
Between the jumps of $N$, the solution $X^\diamond$ moves along the
force field $a$ (solid blue lines).
\label{fig:mj}}
\end{figure}

For simplicity consider a compound Poisson process $N=(N_t)_{t\geq 0}$ defined in \eqref{eq:cpp}.
As in Section \ref{s:is}, consider the coloured jump noise $\dot N^\tau$, being a derivative of the solution of then
Langevin equation with the relaxation time $\tau>0$ driven by $N$:
\begin{equation}
\ddot N^\tau_t=-\frac{1}{\tau}\dot N^\tau_t+\frac{1}{\tau}N_t,\quad N_0^\tau=0.
\end{equation}
Clearly
\begin{equation}
N^\tau_t=\int_0^t(1-\rme^{-\frac{t-s}{\tau}})\,\rmd N_s =\sum_{m=1}^\infty J_m (1-\rme^{-\frac{t-T_m}{\tau}} ) \mathbb{I}_{[T_m,\infty)}(t)
\end{equation}
and
\begin{equation}
\dot N_t^\tau =\frac{1}{\tau}\int_0^t \rme^{-\frac{t-s}{\tau}}\, \rmd  N_s
=\sum_{m=1}^\infty \frac{J_m}{\tau}    \rme^{-\frac{t-T_m}{\tau}}  \mathbb{I}_{[T_m,\infty)}(t).
\end{equation} 
\begin{figure}
\begin{center}
\includegraphics[width=.6\textwidth]{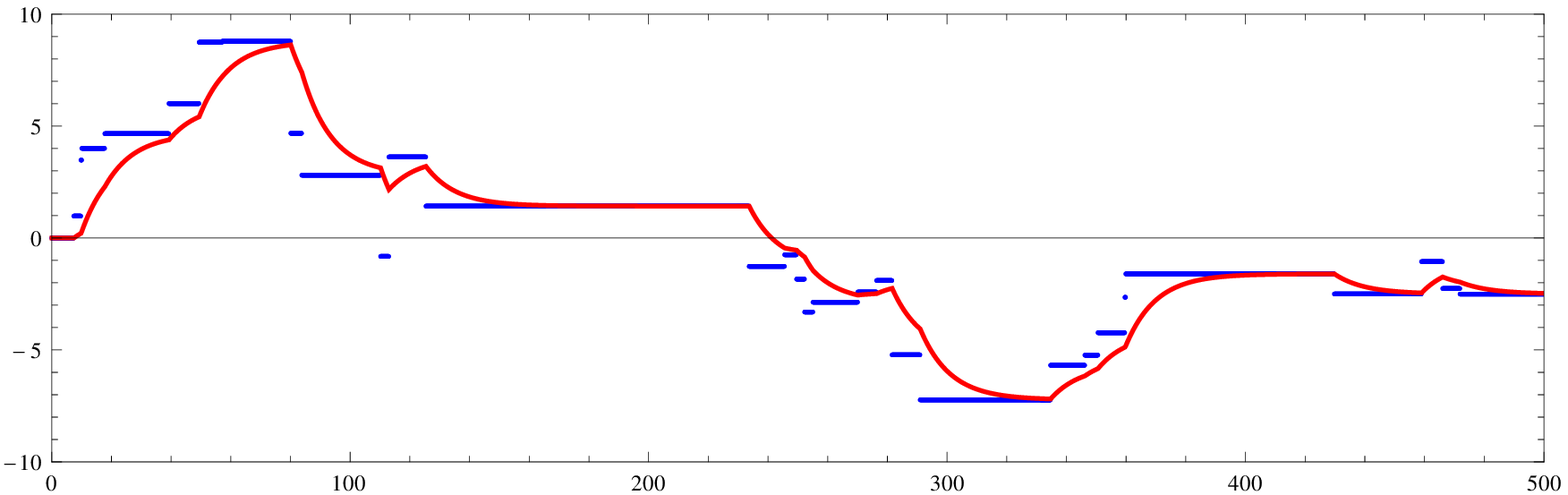}

\includegraphics[width=.6\textwidth]{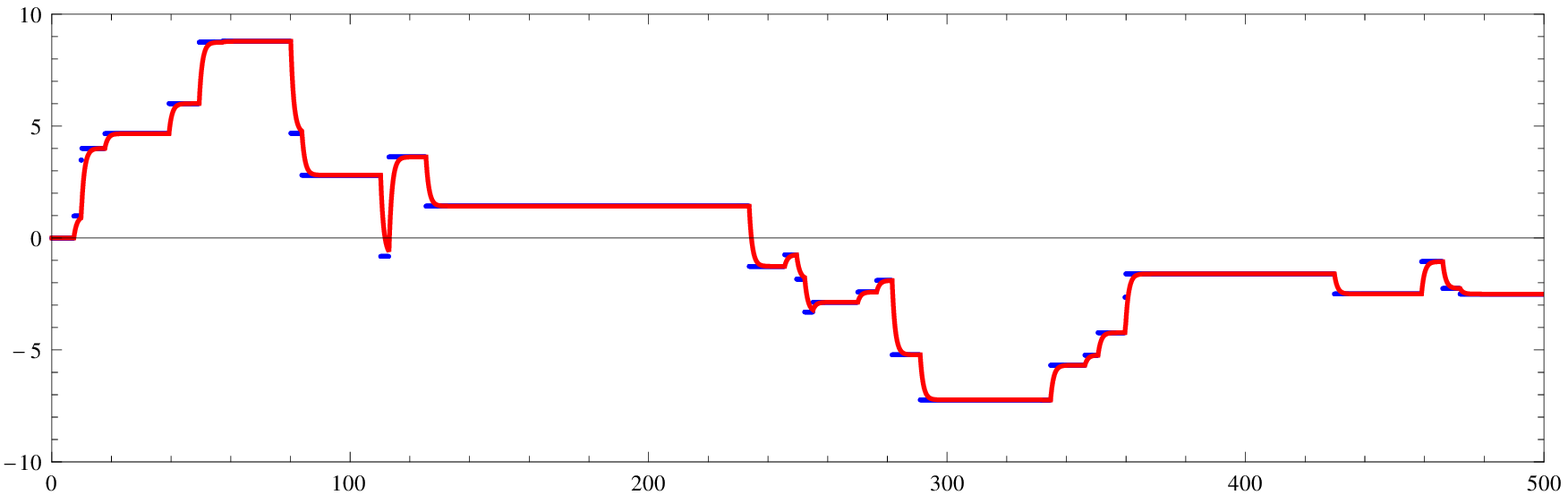}

\includegraphics[width=.6\textwidth]{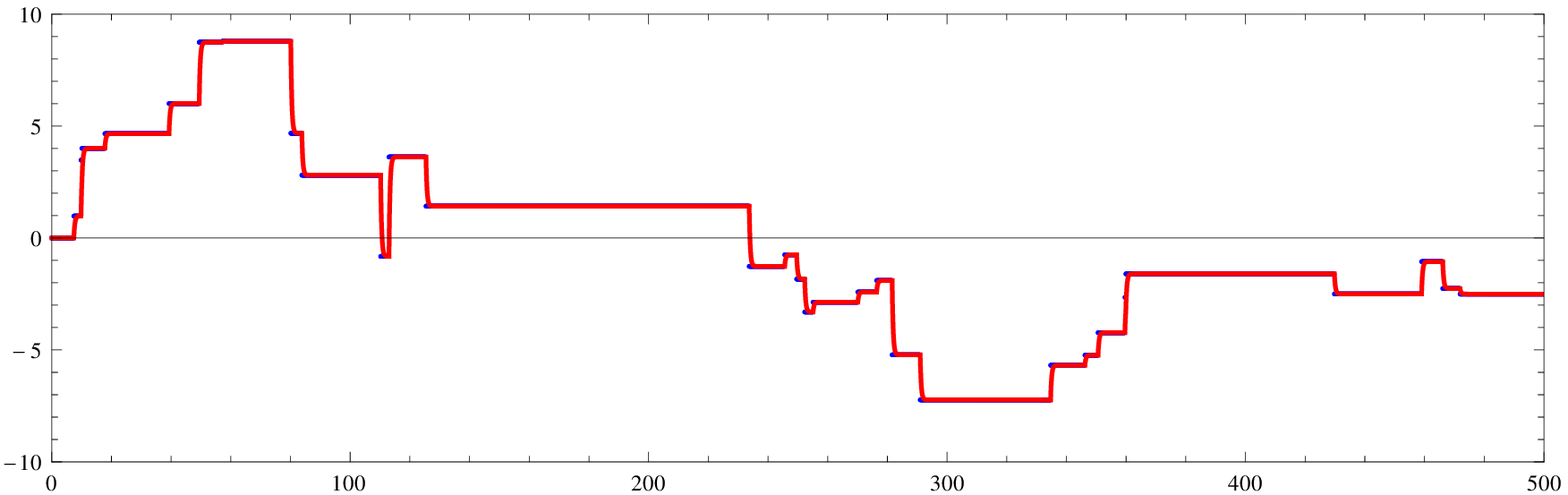}
\end{center}
\caption{(Colour online) Coloured approximations $N^\tau$ (red) of the compound Poisson process $N$ (blue) 
with decreasing relaxation times $\tau=100, 1$, and $0.3$ (from top to bottom).
\label{f:jump}}
\end{figure}
The approximation $N^\tau_t$ converges to $N_t$ as $\tau\to 0$ on the time intervals between the jumps, and monotonically `glues together' 
the discontinuities (see Fig.\ \ref{f:jump}).
For $\tau>0$ consider a random differential equation driven by the multiplicative smoothed process $N^\tau$:
\begin{equation}
X_t^\tau=x+\int_0^t a(X_s^\tau)\,\rmd s+ \int_0^t b(X^\tau_s)\, \rmd N_s^\tau.
\end{equation}
Let us study the limiting behaviour of $X^\tau$ in the limit $\tau \downarrow 0$.
Clearly, between the jumps of $N$ and for small $\tau$, the solution $X^\tau$ moves along the external field $a$. Put for simplicity $a=0$.
Then the equation for $X^\tau$ takes the form
\begin{equation}
\dot X^\tau_t= b(X^\tau_t)\dot N^\tau_t
=b(X^\tau_t)\sum_{m=1}^\infty \frac{J_m}{\tau}\rme^{-\frac{t-T_m}{\tau}} \mathbb{I}_{[T_m,\infty)}(t),
\end{equation}
This is a random non-autonomous differential equation with piece-wise smooth right-hand side. It is natural to solve it sequentially 
on the inter-jump intervals $(T_m, T_{m+1})$. On this time interval the equation has the form
\begin{equation}
\dot X_t^\tau = b(X_t^\tau)\Big[ \sum_{k=1}^{m-1} \frac{J_k}{\tau}    \rme^{-\frac{t-T_k}{\tau}} +
  \frac{J_m}{\tau}    \rme^{-\frac{t-T_m}{\tau}} \Big].
\end{equation}
and the terms $\sum_{k=1}^{m-1} J_k    \rme^{-\frac{t-T_k}{\tau}}$ can be neglected for $\tau$ small
enough such that $\tau \ll \frac{1}{\lambda}=\langle T_m-T_{m-1}\rangle$. 
Then the equation reduces to
\begin{equation}
X_t^\tau =X_{T_m-}^\tau+ \int_{T_m}^{T_m+t} b(X_s^\tau) \frac{J_m}{\tau}    \rme^{-\frac{s-T_m}{\tau}}\, \rmd  s.
\end{equation}
For convenience, we perform the time shift at $T_m$, denote $U_t^\tau=X^\tau_{T_m+t}$ and consider the equation
\begin{equation}
\label{eq:2}
U_t^\tau=X_{T_m-}^\tau+ \int_{0}^{t} b(U_s^\tau) \frac{J_m}{\tau}    \rme^{-\frac{s}{\tau}}\, \rmd  s,\quad t\in[ 0, T_{m+1}-T_m),
\end{equation}
in the limit $\tau\to 0$. To capture the fast change of the solution caused by the jump of $N$ of the size $J_m$ we perform a 
time stretching transformation
\begin{equation}
s=-\tau\ln(1-u),\quad u\in[0,1),\quad
u=1-\rme^{-s/\tau},\quad s\geq 0.
\end{equation}
which transforms  \eqref{eq:2} into
\begin{equation}
\label{eq:1}
U_t^\tau=X_{T_m-}^\tau+ \int_{0}^{1-\rme^{-t/\tau}} b(U_{-\tau\ln(1-u)}^\tau) J_m\, \rmd  u
\end{equation}
Denote
\begin{equation}
Y^\tau_u=U_{-\tau\ln(1-u)}^\tau
\quad \mbox{or equivalently}\quad
U_{t}^\tau=Y^\tau_{1-\rme^{-t/\tau}}. 
\end{equation}
Then \eqref{eq:1} can be rewritten in terms of the process $Y^\tau$ as
\begin{equation}
\label{eq:Y}
Y^\tau_{1-\rme^{-t/\tau}}  =X_{T_m-}^\tau+ \int_{0}^{1-\rme^{-t/\tau}} b(Y_{u}^\tau) J_m\, \rmd  u.
\end{equation}
It is natural to assume that $X^\tau\to X^\diamond$ in the limit $\tau\to 0$. Passing to the limit in equation \eqref{eq:Y} for any $t>0$
we recover the identity
\begin{equation}
\label{eq:id}
Y^0_{1}  =X_{T_m-}^\diamond+ \int_{0}^{1} b(Y_{u}^0) J_m\, \rmd  u.
\end{equation}
The value $Y^0_1$ determines position the of the limiting solution $X^\diamond$ after the jump of the size $J_m$. 
Eq.\ \eqref{eq:id} is the integral form of the ordinary non-linear differential equation 
\begin{equation}
\label{eq:mdiff}
\begin{aligned}
\frac{\rmd }{\rmd u} y(u;x,z)&=b(y(u,x,z))z,\\ 
y(0;x,z)&=x,
\end{aligned}
\end{equation}
with time $u\in[0,1]$, a parameter $z=J_m$ and the initial value $x$ being equal to the value of the solution $X_{T_m-}^\diamond$ 
just before the jump.
Eq.\ \eqref{eq:mdiff} plays a particular role 
in the theory of Marcus equations. 
Indeed, for any $x$ and any $z$ let us denote its solution evaluated at $u=1$ by $\phi^z(x):=y(1;x,z)$.
Then $X^\diamond_{T_m}=\phi^{J_m}(X^\diamond_{T_m}-)$ and the instantaneous jump occurs along the curve 
$y(u; X_{T_m-},J_m)$, $u\in[0,1]$, see Figure \ref{fig:mj}.

Overall, coming back to the process $X^\tau$ and taking into account the drift $a$ we find that in the limit 
$\tau\to 0$ the continuous dynamics of $X^\tau$ obeys the following 
equation with jumps, which is known to be the Marcus (canonical) equation:
\begin{equation}
\label{eq:mo}
X_t^\diamond=x +\int_0^t a(X^\diamond_s)\,\rmd  s +\sum_{m\colon T_m\leq t}\Big(   \phi^{J_m}( X^\diamond_{T_m-})-X^\diamond_{T_m-}\Big).
\end{equation}
Recalling that according to the definition of the It\^o integral we have
\begin{equation}
\label{eq:mark}
\int_0^t b(X_s^\diamond)\, \rmd  N_s=\sum_{s\leq t} b(X_{s-}^\diamond)\Delta N_s=\sum_{m\colon T_m\leq t} b(X^\diamond_{T_m-})J_m,
\end{equation}
we can rewrite \eqref{eq:mo} as an It\^o equation with a correction term
\begin{equation}
\label{eq:m}
X_t^\diamond=x+\int_0^t a(X^\diamond_s)\,\rmd  s  +\int_0^t b(X^\diamond_s)\, \rmd  N_s
+\sum_{m\colon T_m\leq t}\Big(  \phi^{J_m}( X_{T_m-}^\diamond)  -X_{T_m-}^\diamond   -b(X_{T_m-}^\diamond)J_m\Big ).
\end{equation}
The last two terms in the formula \eqref{eq:m} are abbreviated as 
the Marcus `integral' $\int_0^t b(X_s^\diamond)\diamond \rmd  N_s$ with respect to $N$. The equation \eqref{eq:mark}
is thus formally written as
\begin{equation}
\label{eq:mm}
X_t^\diamond=x +\int_0^t a(X^\diamond_s)\,\rmd  s +\int_0^t b(X_s^\diamond)\diamond \rmd  N_s.
\end{equation}
Now it is easy to obtain the stochastic equation for the colour noise limit 
of the dynamics driven by the Brownian motion with drift and a compound Poisson process $N$.
Let $L_t=\omega_0 t+\sigma W_t+ N_t$. Then on the intervals between the jumps of $N$ the solution evolves according to the 
continuous Stratonovich equation and is inter-dispersed with jumps calculated with the help of the mapping $\phi^z(x)$.
Eventually we obtain the equation
\begin{equation}
\label{eq:mmm}
\begin{aligned}
X_t^\diamond=x &+ \int_0^t a(X_s^\diamond)\,\rmd  s+ \int_0^t b(X_s^\diamond)\diamond\rmd  L_s
\\
=x &+ \int_0^t a(X_s^\diamond)\,\rmd  s+ \int_0^t b(X_s^\diamond)\circ\rmd  (\sigma W_s+\omega_0 s )\\
&+\int_0^t b(X^\diamond_s)\, \rmd  N_s
+\sum_{m\colon T_m\leq t}\Big(  \phi^{J_m}( X_{T_m-}^\diamond)  -X_{T_m-}^\diamond   -b(X_{T_m-}^\diamond)J_m\Big )
\end{aligned}
\end{equation}
One can prove (see, e.g. \cite{KurtzPP-95,Kunita-04, Applebaum-09}) that the compound Poisson process $N$ in \eqref{eq:mmm} can be replaced
by a L\'evy process $Z$ with infinitely many jumps, e.g.\ a L\'evy flights process. 
In this case, the Marcus correction term
contains a sum over infinitely many jumps of $Z$ and $X^\diamond$ satisfies
\begin{equation}
\label{eq:mh}
\begin{aligned}
X_t^\diamond
=x &+ \int_0^t a(X_s^\diamond)\,\rmd  s+ \int_0^t b(X_s^\diamond)\circ\rmd  (\sigma W_s+\omega_0 s )\\
&+\int_0^t b(X^\diamond_s)\, \rmd  Z_s
+\sum_{s\leq t}\Big(  \phi^{\Delta Z_s}( X_{s-}^\diamond)  -X_{s-}^\diamond   -b(X_{s-}^\diamond)\Delta Z_s\Big ).
\end{aligned}
\end{equation}

It is clear that for the additive noise, $b(x)\equiv \mbox{Const}$, the Marcus, Stratonovich and It\^o equations coincide.
For multiplicative continuous noise, the Marcus equation coincides with the Stratonovich equation and differs from the It\^o one. 
In the case of multiplicative jump noise, all three equations are different.

Also note that the Marcus `integral' is not an integral but an abbreviation of an It\^o integral and a correction sum from
the last line in \eqref{eq:mmm} or \eqref{eq:mh}. This is why we are not able to calculate
expressions like $\int_0^t P_s\diamond \rmd  P_s$ in the Marcus sense.

Fortunately, the chain rule can be still applied to solutions of Marcus SDEs.
Indeed, for any twice differentiable function $F$
we can write
\begin{equation}
\begin{aligned}
F(X_t^\diamond)
&=F(x)+\int_0^t F'(X_s^\diamond )a(X_s^\diamond )\,\rmd  s+ \int_0^t F'(X_s^\diamond )b(X_s^\diamond )\diamond \rmd  L_s\\
&=F(x)+\int_0^t F'(X_s^\diamond )\diamond X_s^\diamond.
\end{aligned}
\end{equation}
where the term $\int_0^t F'(X_s^\diamond )b(X_s^\diamond )\diamond \rmd  L_s$ can be understood as a small relaxation limit of the
integrals\\ $\int_0^t F'(X_s^\tau )b(X_s^\tau )\, \rmd  L_s^\tau$.

Thus, the Marcus calculus enjoys all the properties one would expect from the Stratonovich integration rule, namely, the conventional 
change of variables formula and the validity of the coloured noise approximations. 

\begin{exa}
Consider an SDE in the sense of Marcus for a real-valued linear system driven by the L\'evy process $L_t=\omega_0 t+\sigma W_t+ N_t$ 
(compare with Example \ref{ex:expjump})
\begin{equation}
X_t^\diamond=1+\int_0^t X_{s}^\diamond\, \diamond \rmd  L_s.
\end{equation}
Since the conventional Newton--Leibniz integration formula applies here, we get
\begin{equation}
X_t^\diamond=\rme^{L_t}=\rme^{\omega_0 t+\sigma W_t+N_t}.
\end{equation}
In particular, in the Poisson case $L=zP$, $z\in\mathbb R$, we obtain (compare with Eqs.\ \eqref{eq:eito} and \eqref{eq:estrat})
\begin{equation}
X_t^\diamond=\rme^{z P_t}.
\end{equation}
\end{exa}

\begin{figure}
\begin{center}
\includegraphics[width=.4\textwidth]{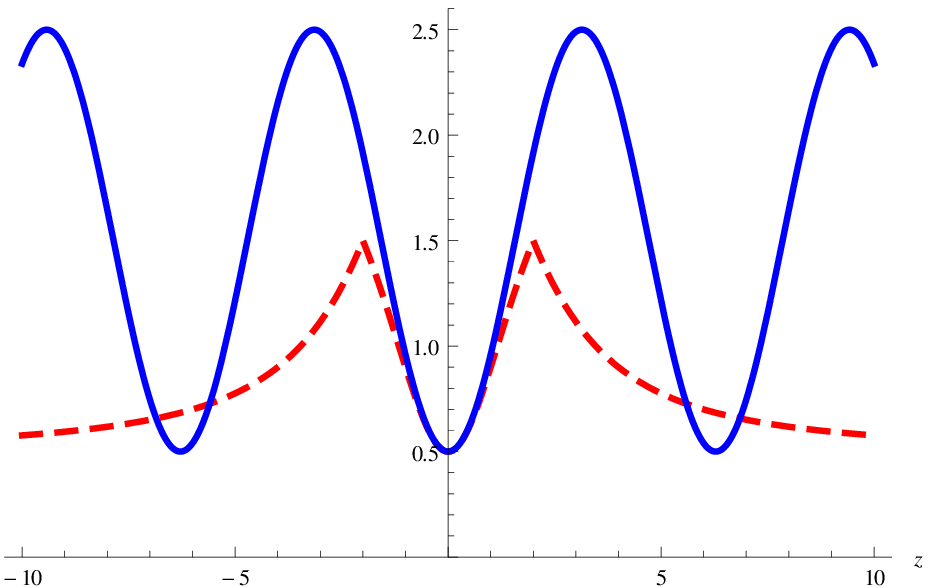}\hspace{1cm}\includegraphics[width=.4\textwidth]{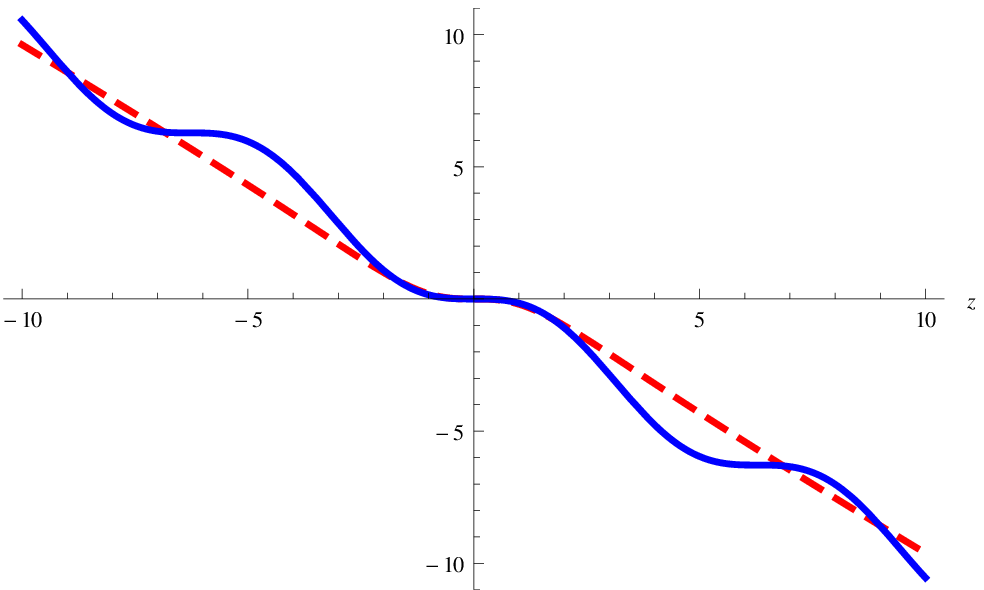}
\end{center}
\caption{(Colour online) The dependence of the spectral line widths $\gamma^\circ$ and $\gamma^\diamond$ (left) 
and phase shifts $\omega^\circ$ and $\omega^\diamond$ (right)
on the jump size $z\in\mathbb R$ for the Stratonovich (red, dashed) and Marcus (blue) Kubo--Anderson oscillators; $\sigma=1$, $\lambda=1$.
\label{f:spectrum}}
\end{figure}

\begin{exa}
Consider the Kubo--Anderson oscillator with Marcus multiplicative noise $l_t=\omega_0+\sigma W_t+z(P_t-\lambda t)$ 
(compare with Example \ref{ex:kubostrat}):
\begin{equation}
Z_t^\diamond=Z_0+\rmi \int_0^t Z_{s}^\diamond\,\diamond \rmd  l_s.
\end{equation}
The solution to this equation is the conventional exponent
\begin{equation}
Z_t^\diamond= Z_0\rme^{\rmi\left(\omega_0 t +\sigma W_t + z(P_t-\lambda t)\right)},
\end{equation}
and this solution is physically meaningful.
As in the Stratonovich case, assume that $|Z_0|=1$ and determine the relaxation function
\begin{equation}
\begin{aligned}
\Phi^\diamond(t)&=\langle \overline{Z_0^\diamond} Z_t^\diamond\rangle 
= \langle\rme^{\rmi\left((\omega_0-\lambda z) t +\sigma W_t + zP_t\right)}\rangle\\
&=\rme^{-t(\frac{\sigma^2}{2} +\lambda(1-\cos z))}\rme^{\rmi t (\omega_0+\lambda\sin z-\lambda z)}=
\rme^{-\gamma^\diamond t+\rmi (\omega_0+\omega^\diamond) t},\\
\gamma^\diamond&=\frac{\sigma^2}{2} +\lambda(1-\cos z),\\
\omega^\diamond&=\lambda(\sin z-z).
\end{aligned}
\end{equation}
Then, similar to the Stratonovich case, Eq.\ \eqref{ex:shapestra}, the line shape is obtained as 
\begin{equation}
\label{ex:shapemar}
I^\diamond(\omega)=\frac{1}{\pi}\mbox{Re}\int_0^\infty \rme^{-\rmi \omega t}\Phi^\diamond(t)\,\rmd  t= 
\frac{1}{\pi}\frac{\gamma^\diamond}{(\gamma^\diamond)^2 +(\omega-\omega_0-\omega^\diamond)^2}.
\end{equation}
The line widths and the frequency shifts in Stratonovich and Marcus cases are shown in Fig. \ref{f:spectrum}
and compared in the next Section.
\end{exa}

\section{Discussion}

For systems with jump noises or bursty fluctuations the Marcus integration plays the same role 
as the Stratonovich 
integration for systems driven by Brownian motion. In this paper, we derived the Stratonovich equation as a small 
correlation time limit of differential equations driven 
by Gaussian Ornstein--Uhlenbeck coloured noise. Analogously, we introduced a Marcus canonical equation as a 
limit of equations driven 
by L\'evy Ornstein--Uhlenbeck coloured noise.
In contrast to the Wong--Zakai polygonal approximation scheme, the Ornstein--Uhlenbeck approximations 
are non-anticipating functions in the sense that they do not account for future events, see \cite[\S  4.2.4]{Gardiner-04}.  
They can be also treated within the theory of Markov processes and Fokker--Planck equation.

We solved explicitly the It\^o, Stratonovich and Marcus equations for two generic linear systems driven by
Brownian motion inter-dispersed by Poisson jumps.
As expected, the Marcus interpretation is consistent with the conventional integration rules. 
The It\^o interpretation of the Kubo--Anderson oscillator demonstrates a physically inappropriate 
solution with exponentially increasing amplitude.
The Stratonovich and Marcus solutions reveal remarkably similar properties. Both solutions do not leave the unit circle on a 
complex plane and have a Lorenzian spectral line shape. However, the frequency shifts and the line widths exhibit 
different behaviour as functions of the jump size $z$.
In particular, in the Marcus case the line width is a periodic function, whereas in the Stratonovich 
case it attains its maxima at $z=\pm 2$ 
and decreases monotonically at larger $|z|$. 

We note that in the theory of Brownian motion, in dependence of the phenomenon considered, 
another important prescription is physically relevant, namely the H\"anggi--Klimontovich prescription, or the so-called 
post-point scheme, see \cite{TikMir-77,Haenggi-78,Haenggi-80a,Klimontovich-90,Klimontovich-94,DunHae-05},
Very recent studies go even beyond the It\^o, Stratonovich or H\"anggi--Klimontovich prescriptions \cite{YuanAo-12,ShiCYYA-12}. 
It would be interesting to extend these approaches also to non-Gaussian jump noises. Another interesting research 
direction would be to give a thermodynamical interpretation 
in case of discontinuous processes. Here we may refer to the two papers \cite{LauLub07,Sokolov-10} on this issue in the theory
of Brownian motion.

\section*{Acknowledgements}

The authors are grateful to the Max Planck Institute for the Physics of Complex Systems, Dresden, Germany, for hospitality.

\newcommand{\etalchar}[1]{$^{#1}$}

%

\end{document}